\documentstyle[multicol,aps,prb,epsfig,floats]{revtex}

\newcommand{\bi}{\begin{itemize}}
\newcommand{\ei}{\end{itemize}}

\newcommand{\be}{\begin{equation}}
\newcommand{\ee}{\end{equation}}

\newcommand{\bea}{\begin{eqnarray}}
\newcommand{\eea}{\end{eqnarray}}
\newcommand{\beastar}{\begin{eqnarray*}}
\newcommand{\eeastar}{\end{eqnarray*}}

\newcommand{\half}{\frac{1}{2}}
\newcommand{\eq}[1]{~(\ref{#1})}
\newcommand{\eqq}[2]{~(\ref{#1},\ref{#2})}
\newcommand{\eqqq}[3]{~(\ref{#1},\ref{#2},\ref{#3})}
\newcommand{\Eq}[1]{ Eq.~(\ref{#1})}
\newcommand{\Eqq}[2]{ Eqs.~(\ref{#1},\ref{#2})}
\newcommand{\Eqqq}[3]{ Eqs.~(\ref{#1},\ref{#2},\ref{#3})}

\newcommand{\order}{{{\mathcal O}}}

\newcommand{\ie}{{\it i.e.}}

\newcommand{\rhol}{\rho(l)}
\newcommand{\rhodl}{\rho_d(l)}
\newcommand{\pll}{P(l)}
\newcommand{\x}{_{\rm x}}
\newcommand{\y}{_{\rm y}}
\newcommand{\z}{_{\rm z}}

\newcommand{\pdl}{P_l(d)}
\newcommand{\pxl}{P_l({\rm x})}
\newcommand{\pyl}{P_l({\rm y})}
\newcommand{\pzl}{P_l({\rm z})}

\newcommand{\rhox}{\rho\x}
\newcommand{\rhoy}{\rho\y}
\newcommand{\rhoz}{\rho\z}
\newcommand{\mom}{\phi}
\newcommand{\dmom}{\Delta}
\newcommand{\intl}{\int\!\!dl\,}
\newcommand{\sd}{\sum_d}
\newcommand{\laml}{\kappa(l)}
\newcommand{\pa}{^{(\al)}}
\newcommand{\pn}{^{(0)}}
\newcommand{\rhoal}{\rho\pa(l)}
\newcommand{\rhon}{\rho^{(0)}}
\newcommand{\pnl}{P^{(0)}(l)}
\newcommand{\rhonl}{\rho\pn(l)}
\newcommand{\rhondl}{\rho_d\pn(l)}
\newcommand{\fmom}{f_{\rm mom}}
\newcommand{\fexc}{\tilde{f}}

\newcommand{\al}{a}
\newcommand{\bet}{{a'}}
\newcommand{\lamb}{\alpha}
\newcommand{\apar}{\lamb_\parallel}
\newcommand{\aper}{\lamb_\perp}
\newcommand{\K}{K}
\newcommand{\I}{^{\rm I}}
\newcommand{\N}{^{\rm N}}
\newcommand{\m}{m}
\newcommand{\rl}{r(l)}

\begin{document}

\title{Phase equilibria in the polydisperse Zwanzig model of hard
rods}

\author{Nigel Clarke$^1$, Jos{\'{e}} A.\ Cuesta$^2$, Richard Sear$^3$,
Peter Sollich$^4$, Alessandro Speranza$^4$}

\address{$^1$Materials
Science Centre, UMIST and The University of Manchester,\\
Manchester M1 7HS, U.K.; {\tt nigel.clarke@umist.ac.uk}
\\
$^2$GISC,
Dpto.\ de Matem{\'{a}}ticas, Universidad Carlos III de Madrid,
Avda.\ de la Universidad 30\\
28911 - Legan{\'{e}}s, Madrid, Spain;
{\tt cuesta@math.uc3m.es}
\\
$^3$Department of Physics, University of Surrey, Guildford,\\
Surrey GU2 7XH, U.K.; {\tt R.Sear@surrey.ac.uk}
\\
$^4$Department of Mathematics, King's College London, Strand,\\
London WC2R 2LS, U.K.;
{\tt \{peter.sollich,alessandro.speranza\}@kcl.ac.uk}
}

\maketitle

\begin{abstract}
We study the phase behaviour of the Zwanzig model of suspensions of
hard rods, allowing for polydispersity in the lengths of the rods. In
spite of the simplified nature of the model (rods are restricted to
lie along one of three orthogonal axes), the results agree
qualitatively with experimental observations: the coexistence region
broadens significantly as the polydispersity increases, and strong
fractionation occurs, with long rods found preferentially in the
nematic phase. These conclusions are obtained from an analysis of the
exact phase equilibrium equations. In the second part of the paper, we
consider the application of the recently developed ``moment free
energy method'' to the polydisperse Zwanzig model. Even though the
model contains non-conserved densities due to the orientational
degrees of freedom, most of the exactness statements (regarding the
onset of phase coexistence, spinodals, and critical points) derived
previously for systems with conserved densities remain valid. The
accuracy of the results from the moment free energy increases as more
and more additional moments are retained in the description. We show
how this increase in accuracy can be monitored without relying on
knowledge of the exact results, and discuss an adaptive technique for
choosing the extra moments optimally.
\end{abstract}

\begin{multicols}{2}

\section{Introduction}

Solutions of rod-like particles undergo a transition from a disordered
isotropic phase to an ordered nematic phase as the concentration of
rods is increased. In the isotropic phase the rods have no preferred
orientation, whereas in the nematic phase there is a favoured average
alignment of the rods. Such a transition has been observed by
Zocher~\cite{zocher25} in solutions of rod-like $V_{2}O_{5}$
particles, and later by Bawden, Pirie, Bernal and
Fankuchen~\cite{bawden36,bernal41} in solutions of tobacco mosaic
virus. In 1949, Onsager~\cite{onsager49} showed that the ordering
could be explained, at least for solutions of monodisperse long thin
rigid rods, by considering only the competition between the excluded
volume interaction and the orientational entropy. The key to his model
is the distribution function, $P(\theta),$ that represents the
fraction of rods with a given angular orientation, $\theta$, with
respect to some director. In principle, this function can be
determined self-consistently by minimizing the free energy; however,
the resultant non-linear integral equation for $P(\theta)$ cannot be
solved analytically. To progress further, Onsager introduced a trial
function with a single variational parameter, $\alpha$. By now
minimizing the resultant free energy with respect to $\alpha$, its
concentration dependence can be calculated. From this dependence the
conditions for coexistence of the isotropic and nematic phases can be
determined.

There are several difficulties in quantitatively comparing
experimental results, from systems such as solutions of tobacco mosaic
virus, with Onsager's predictions. Firstly, in addition to the hard
core repulsion, the interaction between rods often includes a soft
repulsion due to electrostatic forces. Secondly, the rods are rarely
rigid, and semi-flexibility must be accounted for. To overcome these
difficulties, Buining, Veldhuizen, Pathmamanoharan, Hansen and
Lekkerkerker~\cite{buining91,buining92} synthesized sterically
stabilized beohmite particles. Buining and
Lekkerkerker~\cite{buining93} and van Bruggen, van der Kooij and
Lekkerkerker~\cite{vanbruggen96} then studied the phase behaviour of
solutions of these particles, which can be modelled reasonably with a
rigid hard rod interaction, with corrections arising from soft
electrostatic repulsions and semi-flexibility being
negligible. However, there remains one further complication: as with
most polymeric systems, the particles are polydisperse.

The Onsager approach has been modified to enable phase diagrams for
bidisperse~\cite{odijk85,lekkerkerker84,vroege93} and
tridisperse~\cite{vroege97} systems, in which the rods differ only in
length, to be calculated. A rich variety of behaviour has been
predicted, such as widening of the region of coexistence,
fractionation of the longer rods into the nematic phase,
nematic-nematic coexistence, and re-entrant phases. Qualitatively, the
first three of these are in agreement with the experiments reported in
Ref.~\onlinecite{buining93} and~\onlinecite{vanbruggen96}, and the widening
of the region of coexistence between phases appears to be a general
feature of polydisperse systems. Despite the obvious difficulties in
trying to map a continuous polydisperse system onto a bidisperse
mixture, Merchant and Rill~\cite{merchant97} have attempted to analyse
the transition concentration in solutions of polydisperse rod-like DNA
in terms of the theory presented in
Ref.~\onlinecite{lekkerkerker84}. It is clear that quantitative
agreement is still lacking. The generalisation of the Onsager approach
to continuous polydispersity is, however, far from simple. The only
attempts, to date, that
\end{multicols}
\twocolumn
we are aware of, have treated the polydispersity
perturbatively~\cite{Sluckin89,chen94}; this limits the validity of
the analysis to situations with rather narrow distributions of rod
lengths.

Another approach to the problem of monodisperse rod-like mixtures was
adopted by Zwanzig~\cite{zwanzig63}: he restricted the orientations of
the rods to be in one of three mutually perpendicular directions. This
enables the exact calculation of higher order virial coefficients, and
the orientational distribution can be determined without
approximations. In contrast to the Onsager approach, the Zwanzig model
may be readily extended to polydisperse systems. For bidisperse
mixtures it has already been shown by Clarke and
McLeish~\cite{clarke92} that the qualitative features of the phase
diagram, with the exception of the nematic-nematic coexistence, are
similar to those predicted by Lekkerkerker, Coulon, van der Haegen and
Debliek~\cite{lekkerkerker84}. The polydisperse Zwanzig model
therefore provides a useful starting point for understanding the
effects of polydispersity on the phase behaviour of hard rod systems.

A further important motivation for studying polydispersity, in this
simplified model, is that it provides an interesting scenario for
testing and extending the recently proposed ``moment method'' approach
to the thermodynamic treatment of polydisperse
systems~\cite{SolCat98,Warren98,polydisp_long}. The moment method
applies to systems whose excess free energy depends only on some {\em
moments} of the density distribution describing a polydisperse system;
we show below that the Zwanzig model (treated within the second virial
approximation) is of exactly this form. By expressing the ideal part
of the free energy in a similar form a ``moment free energy'' can be
defined, which only depends on the given moment densities. This is a
drastic reduction in the number of densities required to describe the
system, from the infinite number of degrees of freedom of the complete
density distribution to a finite number of moment densities. The
standard methods of the thermodynamics of finite mixtures can then be
applied to the moment free energy to analyse the phase
behaviour. Although in general the results will be approximate, for
systems with conserved densities it has been
shown~\cite{SolCat98,Warren98,polydisp_long} that the cloud and shadow
points (which specify the onset of phase coexistence in polydisperse
systems), spinodals, and critical points are all found {\em
exactly}. In the case of the Zwanzig (or Onsager) model, however, one
has both conserved (rod lengths) and non-conserved (rod orientations)
degrees of freedom. We show below that the moment method can be
extended to this kind of scenario, and that most of the above
exactness statements carry over. We also assess the accuracy of the
moment method in the region where it is not exact and discuss how it
can be improved systematically by retaining additional moment
densities in the description.


The paper is structured as follows. In Sec.~\ref{sec:model}, we
describe the polydisperse Zwanzig model, give its free energy
(calculated within the second virial approximation) and derive the
corresponding expressions for the chemical potentials and the osmotic
pressure. These results are used in Sec.~\ref{sec:exact} to study the
exact thermodynamic behaviour of the model. In Sec.~\ref{sec:moment},
we construct the moment free energy, discuss its properties, and
compare the results obtained from it with the exact ones. Because the
exact calculation of the phase behaviour of the polydisperse Zwanzig
model is feasible, it might seem unnecessary to confirm these results
using the moment method. However, it is precisely {\em because} the
exact results are available that the Zwanzig model provides a useful
test case for the application of the moment method to systems with
non-conserved degrees of freedom. The conclusions drawn should help us
to apply the method to more complicated systems (such as the
polydisperse Onsager model) where an exact calculation of the phase
behaviour is infeasible. This and other possible avenues for future
work are discussed in Sec.~\ref{sec:conclusion}.

\section{Definition of the model}
\label{sec:model}

The Zwanzig model considers hard rods in the shape of parallelepipeds
of length $L$ and with square base of edge length $D$. In contrast to
the full Onsager model, the orientations of these rods are restricted
to be along one of the three cartesian coordinate axes, x, y or
z. We will assume that all rods have the same diameter $D$, but that
they are polydisperse in {\em length}, so that there is a continuous
distribution of rod lengths $L$. Introducing a reference length $L_0$
and the normalized lengths $l=L/L_0$, we will focus on the Onsager
limit of long thin rods. This corresponds to letting $D/L_0\to 0$
while keeping the normalized lengths $l$ constant. Unless a
distinction between normalized ($l$) and unnormalized ($L$) lengths
needs to be made explicitly, we will simply refer to $l$ as the length
of a rod in the following.

Because of the length polydispersity, the number densities of the rods
in the three possible orientations are specified by density {\em
distributions} $\rhodl$, with $d=$ x, y, z; for a small range of lengths
$dl$, $\rhodl\, dl$ is the number density of rods oriented along $d$
and with lengths between $l$ and $l+dl$. The total number density
distribution (irrespective of orientation) is then
\[
\rho(l)= \rhox(l)+\rhoy(l)+\rhoz(l)
\equiv \sum_d \rhodl
\]
and integrating over $l$ gives the total number density of rods:
\be
\rho = \intl \rhol = \sd \intl \rhodl
\label{rho_def}
\ee
Here and in the following, all integrals over $l$ run from 0 to
$\infty$. With these definitions, the density distribution $\rhodl$
over lengths $l$ and orientations $d$ can be decomposed as
\be
\rhodl = \rhol \pdl = \rho \pll \pdl
\label{rho_decomp}
\ee
where
\[
\pll = \rhol/\rho
\]
is the normalized distribution of rod lengths $l$, and
\[
\pdl = \frac{\rhodl}{\rhol}
\]
is the probability of finding a rod with {\em given} length $l$ in
orientation $d$. Note that $\pdl$ is the analogue of the orientation
distribution in the full Onsager model and obeys the normalization
\be
\sum_d \pdl = 1
\label{pd_norm}
\ee
In the isotropic phase, $\pdl = 1/3$ for all $d$ and $l$. In the
nematic phase, on the other hand, we have $\pxl=\pyl<\pzl$ if we
take the director to be along the $z$-axis. We will nevertheless
develop the theory of the model first for arbitrary orientation
distributions (with $\pxl\neq\pyl$) because this leads to somewhat
more compact expressions, and only specialize to the nematic case
at a later stage.

For simplicity, we only treat the model in the second virial
approximation. (In contrast to the case of the Onsager model, this
approximation does not become exact in the limit $D/L_0\to 0$
here: higher order virial terms to not vanish~\cite{zwanzig63}.)
The excess free energy is then essentially determined by the
excluded volume of two rods. If the rods have (normalized) lengths
$l$ and $l'$ and are {\em perpendicular}, this volume is
\[
V_{\perp}^{\rm excl}=2D(L+D)(L'+D)=2ll'(DL_0^2)[1+\order(D/L_0)]
\]
while the excluded volume for {\em parallel} rods
\[
V_{\parallel}^{\rm excl}=4D^2(L+L')=4(l+l')(DL_0^2)\frac{D}{L_0}=
DL_0^2 \,\order(D/L_0)
\]
is negligible by comparison in the limit $D/L_0\to 0$. As suggested by
the result for $V_\perp^{\rm excl}$, we choose $DL_0^2$ as our unit of
volume in the following, making all densities dimensionless
($\rho\to\rho DL_0^2$). If we also set $k_{\rm B}T=1$, the excess free
energy density becomes (within the second virial approximation)
\bea
\fexc &=& 2\intl dl'\,
ll' [\rhox(l)\rhoy(l') + \rhox(l)\rhoz(l') + \rhoy(l)\rhoz(l')] \nonumber\\
&=& 2(\mom\x\mom\y+\mom\x\mom\z+\mom\y\mom\z) \nonumber\\
&=& \sd\mom_d(\mom-\mom_d)
\label{fexc}
\eea
where we have defined
\begin{mathletters}
\label{phi_def}
\bea
\mom_d&=&\intl l\rhodl=\intl l\rhol\pdl
\label{phid_def}
\\
\mom &=& \sd \mom_d = \intl l \rhol
\eea
\end{mathletters}
Our choice of volume units implies that $(D/L_0)\mom_d$ and
$(D/L_0)\mom$ are, respectively, the volume fraction of rods pointing
in direction $d$ and the total volume fraction. The ratio
\[
\m=\mom/\rho = \intl l \pll
\]
is then simply the average length of rods in the system, \ie, the
first moment of the rod length distribution $\pll$.

Adding the ideal part of the free energy (density) to\Eq{fexc}, we
have for the total free energy (density)
\be
f = \sd\intl\rhodl[\ln\rhodl-1]+\sd\mom_d(\mom-\mom_d)
\label{free_en}
\ee
This equation is the starting point of our analysis. When we refer
to ``exact'' results in the following, we mean the exact
thermodynamics of the model defined by the free
energy\eq{free_en}.

Eq.~(\ref{free_en}) shows that the free energy is a functional of the
density distribution $\rhodl$ over $l$ and $d$. If we
use\Eq{rho_decomp} to write $\rhodl=\rhol\pdl$, we note an important
difference between the two factors: while the total density
distribution $\rhol$ is conserved (because the rods cannot change
length), the orientation distribution $\pdl$ is not (because the rods
can change orientation). We separate out the respective
contributions to the free energy by writing
\bea
f &=& \intl \rhol[\ln\rhol-1]+\intl \rhol \sd\pdl\ln\pdl
\nonumber\\
 & & {} + {} \sd\mom_d(\mom-\mom_d)
\label{free_en_d}
\eea
For a given $\rhol$, the orientation distributions $\pdl$ (for
each $l$) are then obtained by minimizing $f$, subject to the
normalization constraints\eq{pd_norm} (again, for each $l$).
Introducing Lagrange multipliers $\laml$ for these constraints
gives the minimization condition
\beastar
& & \frac{\delta}{\delta \pdl} \left(f+\intl \laml\sd\pdl\right) = 
\nonumber\\
& & \rhol[\ln\pdl+1]+2(\mom-\mom_d)l\rhol + \laml = 0
\eeastar
Solving for $\pdl$ and eliminating the $\laml$ by using\Eq{pd_norm} gives
the orientation distributions
\be
\pdl = \frac{e^{2(\mom_d-\mom)l}}{\sum_{d'}e^{2(\mom_{d'}-\mom)l}}
= \frac{e^{2\mom_dl}}{\sum_{d'}e^{2\mom_{d'}l}}
\label{pdl}
\ee
Inserting this result into the definition\eq{phid_def} then gives
three simultaneous nonlinear equations which can be solved for the
$\mom_d$.

To derive the conditions for phase coexistence in the polydisperse
Zwanzig model, we need expressions for the chemical potential
$\mu(l)$---which, due to the polydispersity, is a function of the
rod length $l$---and the osmotic pressure. The chemical potential
\[
\mu(l)=\frac{\delta f}{\delta \rhol}
\]
is obtained by functional differentiation of the free
energy\eq{free_en_d} w.r.t.\ $\rhol$. There is no contribution
from the variation of $\pdl$ with $\rhol$ because we have already
minimized the free energy w.r.t. $\pdl$. This leads to
\be
\mu(l) =\ln\rhol + \sd \pdl \ln \pdl + 2\sd (\mom-\mom_d)\pdl l
\label{chem_pot_first}
\ee
or, using\Eq{pdl},
\be
\mu(l) = \ln\rhol - \ln\left(\sum_{d}e^{2(\mom_{d}-\mom)l}\right)
\label{chem_pot}
\ee
The osmotic pressure can be written in terms of the free energy
and the chemical potential; hence, using\Eq{chem_pot_first},
\be
\Pi = -f + \intl \rhol \mu(l) = \rho +
\sd \mom_d(\mom-\phi_d)
\label{osm}
\ee

\section{Exact phase coexistence calculation}
\label{sec:exact}

\subsection{Coexistence conditions}
\label{sec:exact_coex_cond}

We can now state the conditions for coexistence of two or more
phases, labelled by $\al=1\ldots \K$, into which a ``parent''
phase with density distribution $\rhonl$ is assumed to have split.
From\Eq{chem_pot}, the equality of the chemical potentials between
the phases is obeyed exactly if the densities can be written in
the form
\be
\rhoal = R(l) \sd \exp\left[\lamb_d\pa l\right]
\label{exact_family}
\ee
with a function $R(l)$ common to all phases and the $\lamb_d\pa$
obeying
\be
\lamb_d\pa = 2 (\mom_d\pa-\mom\pa)+c
\label{lambda_cond}
\ee
Here $c$ is an arbitrary constant (again common to all phases).  If
the phases occupy fractions $v\pa$ of the total system volume,
particle conservation implies
\be
\sum_\al v\pa\rhoal = \rhonl
\label{lever}
\ee
This fixes $R(l)$, giving
\[
\rhoal = \rhonl\, \frac{\sd \exp\left[\lamb_d\pa l\right]}{\sum_\bet
v^{(\bet)} \sum_{d'} \exp\left[\lamb_{d'}^{(\bet)} l\right]}
\]
The density distributions over rod lengths $l$ {\em and} orientations
$d$ are then found from\Eq{rho_decomp} and,
using\Eqq{pdl}{lambda_cond}, take the simple form
\be
\rho_d\pa(l) = \rhonl\, \frac{\exp\left[\lamb_d\pa l\right]}{\sum_\bet
v^{(\bet)} \sum_{d'} \exp\left[\lamb_{d'}^{(\bet)} l\right]}
\label{general_rhodl}
\ee
Integrals over these distributions define, by\Eqq{rho_def}{phi_def},
the values of the densities $\rho\pa$ and volume fractions
$\mom_d\pa$, $\mom\pa$ in all phases. These variables determine the
pressures
\be
\Pi\pa = \rho\pa + \sd \mom_d\pa(\mom\pa-\mom\pa_d)
\label{osm_cond}
\ee
in the different phases; at phase coexistence, these must of course
all be equal. We thus have, in the most general form of the conditions
for coexistence of $\K$ phases, $4\K$ variables (three $\lamb_d\pa$
and one $v\pa$ per phase $\al=1\ldots \K$) and equally many equations
to solve: the $3\K$ conditions\eq{lambda_cond} for chemical potential
equality, the $\K-1$ conditions\eq{osm_cond} for equality of the
pressures, and the trivial normalization of the phase
volume fractions, $\sum_\al v\pa=1$.

It is easy to show that, just as in the Onsager model,
isotropic-isotropic coexistence is not possible in the Zwanzig model
with length polydispersity only. (This would be different if the rod
{\em diameters} were polydisperse as well; compare
Refs.~\onlinecite{SeaJac95,SeaMul96,VanMul96b,VanMulDij98}.)  Given the
results for the bidisperse case~\cite{clarke92}, it is also unlikely
that nematic-nematic coexistence could occur; this is in contrast to
what has been found for the Onsager
model~\cite{lekkerkerker84}. Intuitively, the difference can be
explained as follows: when a polydisperse nematic phase splits into
two nematics containing predominantly short and long rods,
respectively, it gives up entropy of mixing but gains orientational
entropy. In the Onsager model, where the rod angles are continuous
variables, the gain in orientational entropy can be arbitrarily large,
thus favouring such a phase split. (The orientational entropy tends to
$-\infty$ as the orientational distribution function tends to a delta
function.)  In the Zwanzig case, on the other hand, the maximum gain
in orientational entropy is $k_{\rm B}\ln 3$ (this being the
difference between the entropies of an isotropic and a fully ordered
nematic phase) so that nematic-nematic coexistence is disfavoured.

We therefore now specialize to coexistence between an isotropic (I)
and a nematic (N) phase. If we choose the director to be along the
$z$-axis, we then have $\mom\x=\mom\y$ and
$\mom=2\mom\x+\mom\z$. Denoting
\[
\dmom = \mom\z-\mom\x = \mom\z-\mom\y
\]
the volume fractions of rods with the three possible orientations can
be expressed as
\be
\mom\x=\mom\y=\frac{1}{3}(\mom-\dmom), \qquad
\mom\z=\frac{1}{3}(\mom+2\dmom)
\label{mom_simpl}
\ee
and the excess free energy and pressure simplify to
\bea
\fexc &=& \frac{2}{3}(\mom^2-\dmom^2)
\label{fexc_nematic}
\\
\Pi &=& \rho + \frac{2}{3}(\mom^2-\dmom^2)
\label{osm_nematic}
\eea
Instead of numbering the phases by $\al=1,2$, we label them with
superscripts I and N from now on. In the isotropic phase, we have
$\mom\I_d=\mom\I/3$ for $d=$ x,y,z ($\dmom\I=0$), and by choosing the
arbitary constant $c$ in\Eq{lambda_cond} as $c=4\mom\I/3$ we can
ensure that all the coefficients $\lamb\I_d$ vanish. If, for the
corresponding coefficients in the nematic phase, we write
\[
\lamb\N\x=\lamb\N\y\equiv\aper, \qquad
\lamb\N\z\equiv\apar
\]
the conditions\eq{lambda_cond} simplify to
\bea
\apar &=& \frac{4}{3} (\mom\I-\mom\N +\dmom)
\label{apar}
\\
\aper &=& \frac{4}{3} (\mom\I-\mom\N-\half\dmom)
\label{aper}
\eea
The condition\eq{osm_cond} of equality of the pressures, on the other
hand, becomes
\be
\rho\I + \frac{2}{3}(\mom\I)^2 = \rho\N + \frac{2}{3}[(\mom\N)^2-\dmom^2]
\label{osm_IN}
\ee
Note that we have dropped the subscript `N' on $\apar$, $\aper$ and
$\dmom$ because the corresponding quantities in the isotropic phase
are all zero.

If we denote $v\I=\gamma$ and $v\N=1-\gamma$ (so that $\gamma$ is the
fraction of the system volume occupied by the isotropic phase), our
phase coexistence problem now takes the form of three nonlinear
equations\eqqq{apar}{aper}{osm_IN} for $\apar$, $\aper$ and $\gamma$.
The densities and volume fractions appearing in these equations can be
found by specializing\Eqqq{rho_def}{phi_def}{general_rhodl} to the
case of I-N coexistence:
\begin{mathletters}
\label{IN_densities}
\bea
\rho\I &=& \intl \rho\I(l)  \\
\mom\I &=& \intl l\rho\I(l) \\
\rho\N &=& \intl \rho\N(l)  \\
\mom\N &=& \intl l\rho\N(l) \\
\dmom  &=& \intl l[\rho\N_\parallel(l)-\rho\N_\perp(l)]
\eea
\end{mathletters}
with
\begin{mathletters}
\label{IN_distributions}
\bea
\rho\I(l) &=& \frac{3\rhonl}{3\gamma+(1-\gamma)\left[
e^{\apar l} + 2e^{\aper l}\right]}
\\
\rho\N_\parallel(l) &=& \frac{1}{3}\rho\I(l)e^{\apar l}
\\
\rho\N_\perp(l) &=& \frac{1}{3}\rho\I(l)e^{\aper l}
\\
\rho\N(l) &=& \rho\N_\parallel(l) + 2\rho\N_\perp(l)
\eea
\end{mathletters}

In the above setting of the phase coexistence problem, we specified a
single parent density distribution $\rhonl$. We will normally be
interested in results along a so-called ``dilution line'', where the
parent length distribution $\pnl=\rhonl/\rhon$ is kept fixed while the
overall parent density $\rhon$ is varied. As $\rhon$ is increased from
zero, we then expect to find a single isotropic phase first
($\gamma=1$). At the isotropic ``cloud point'', an infinitesimal
fraction of nematic phase will first appear; the density $\rho\N$ of
this nematic phase gives the nematic ``shadow''. On the other hand,
starting from high density $\rhon$ we will first see a pure nematic
phase ($\gamma=0$). On decreasing $\rhon$, an infinitesimal amount of
isotropic phase will appear at the nematic cloud point; the density of
the isotropic phase at this point gives the corresponding isotropic
shadow. The two cloud points delimit the coexistence region. For
values of $\rhon$ inside this region, an isotropic and a nematic phase
coexist and occupy non-infinitesimal fractions of system volume, with
$\gamma$ decreasing from 1 to 0 as $\rhon$ increases.

Numerically, rather than changing $\rhon$ and finding $\gamma$, it
is easier to vary $\gamma$ between 0 and 1 and find the
corresponding $\rhon$. To implement this scheme, one only has to
replace $\rhonl$ in Eqs.\eq{IN_distributions} by $\rhonl=\rhon\pnl$ and
solve\Eqqq{apar}{aper}{osm_IN} for $\apar$, $\aper$ and $\rhon$.
Alternatively, one can interpret $\apar$ and $\aper$ as being
defined by\Eqq{apar}{aper}, and regard $\rhon$, $\rho\I$,
$\mom\I$, $\rho\N$, $\mom\N$ and $\dmom$ as the underlying
variables. Eqs.~(\ref{osm_IN},\ref{IN_densities}) then constitute six
equations for these six unknowns, which can be solved numerically;
this is the approach that we adopt.

\subsection{Results: Cloud point and shadow curves}

In the following, we will restrict ourselves to the case where the rod
lengths in the parent phase are distributed according to a Schulz
distribution
\begin{equation}
\pnl=\frac{(z+1)^{z+1}}{\Gamma(z+1)}\ l^z\exp\left[-(z+1)l\right]
\label{schulz}
\end{equation}
This distribution is normalized and has an average rod length of
$\m\pn=1$. (Allowing other values of $\m\pn$ would not make our
treatment more general since the value of $\m\pn$ can always be
absorbed into a rescaling of the reference length $L_0$.)  The
parameter $z$ controls the shape and width of the distribution,
and is taken to be nonnegative. A more intuitive measure of the
width of the parent distribution is the relative standard
deviation $\sigma$ (usually called the ``polydispersity''),
defined by
\begin{equation}
\sigma^2=\left(\frac{1}{[\m\pn]^2}\intl l^2 \pnl\right) -1
\label{sigma}
\end{equation}
It is then easy to see that, for the Schulz distribution,
\[
\sigma = (1+z)^{-1/2}
\]
For $z\to\infty$, we thus have a monodisperse parent with $\sigma=0$
and $\pnl=\delta(l-1)$. As $z$ is decreased, $\sigma$ increases and
the parent gets more and more polydisperse. For $z=0$, finally, the
parent distribution is a simple exponential, and $\sigma$ achieves its
(for the chosen Schulz distribution) maximal value of 1.

To calculate the cloud point and shadow curves, we proceed as
explained in Sec.~\ref{sec:exact_coex_cond}. For the isotropic cloud
point and shadow, we set $\gamma=1$. In Eqs.\eq{IN_distributions}, we
then have $\rho\I(l)=\rhonl$. This of course makes sense: only an
infinitesimal amount of nematic phase has appeared, and so the density
distribution of the isotropic phase is only negligibly perturbed away
from the parent. In Eqs.\eq{IN_densities}, the equations for $\rho\I$
and $\mom\I$ then simplify to the trivial statements $\rho\I=\rhon$
and $\mom\I=\rhon\m\pn=\rhon$, and we only have to solve four
equations for the four unknowns $\rhon$, $\rho\N$, $\mom\N$ and
$\dmom\N$. Conversely, for the nematic cloud point and shadow, we set
$\gamma=0$. We then find $\rho\N(l)=\rhonl$ and $\rho\N=\mom\N=\rhon$
and have to solve the remaining four equations for the four unknowns
$\rhon$, $\rho\I$, $\mom\I$ and $\dmom\N$.

\begin{figure}
\begin{center}
\epsfig{file=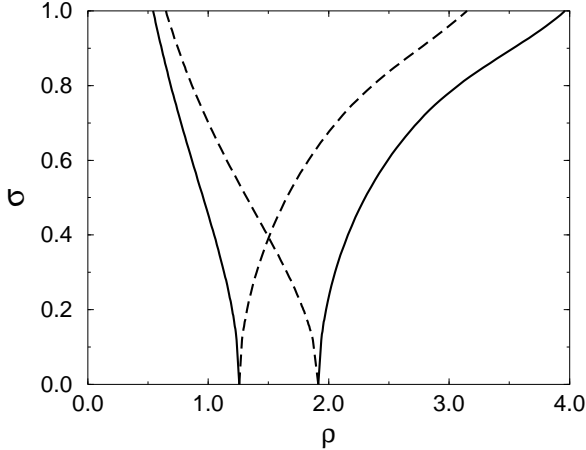,width=8.5cm}
\end{center}
\caption{The isotropic and nematic cloud point curves (solid) and
their corresponding shadow curves (dashed), showing the densities of
the coexisting phases as a function of the polydispersity $\sigma$. Of
the two cloud point curves, the isotropic one is the one with the
lower density; it meets the isotropic shadow curve
for $\sigma\rightarrow0$ as it must. The nematic cloud and
shadow curves likewise coincide in this limit.
\label{fig1}
}
\end{figure}
\begin{figure}
\begin{center}
\epsfig{file=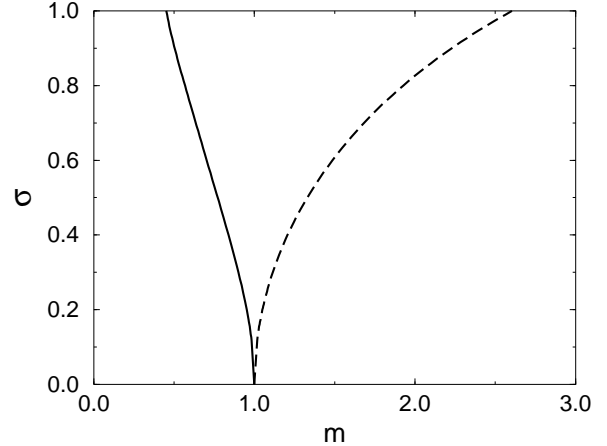,width=8.5cm}
\end{center}
\caption{The average rod lengths $\m$ in the shadow phases are plotted
as a function of the polydispersity, $\sigma$. The dashed and solid
curves give the results for the nematic and isotropic shadow phases,
respectively. Note that the corresonding cloud phases are identical to
the parent and therefore have average rod lengths equal to $\m\pn=1$.
\label{fig2a}
}
\end{figure}
\begin{figure}
\begin{center}
\epsfig{file=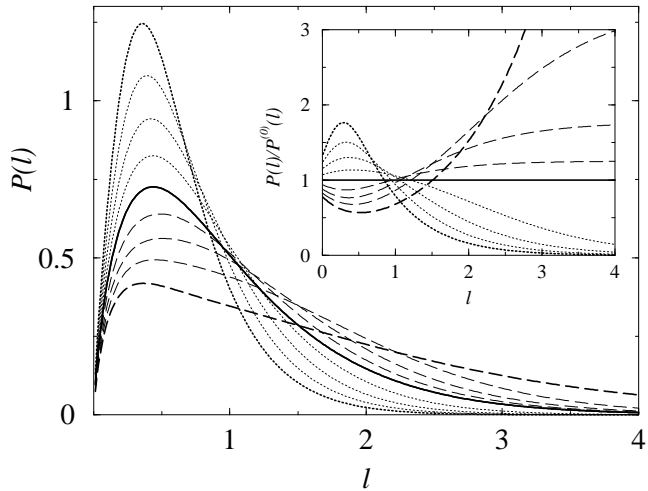,width=8.5cm}
\end{center}
\caption{The rod length distributions $\pll$ in the coexisting
isotropic and nematic phases, at polydispersity $\sigma=0.75$ and for
different fractions $\gamma$ of system volume occupied by the
isotropic phase. Bold dashed line: $\pll$ in the nematic shadow at the
isotropic cloud point, $\gamma=1$. The distribution in the isotropic
phase at this point is the parental one (bold solid line). As $\gamma$
decreases through $0.75, 0.5, 0.25$, the dashed and dotted lines
show---from bottom to top---the distributions in the coexisting
nematic and isotropic phases, respectively. At $\gamma=0$, finally
(the nematic cloud point), the nematic has the parental length
distribution; the bold dotted line shows $\pll$ in the corresponding
isotropic shadow. Inset: Ratio of the rod length distributions to that
of the parent.
\label{fig2b}
\label{gamdist}
}
\end{figure}
\begin{figure}
\begin{center}
\epsfig{file=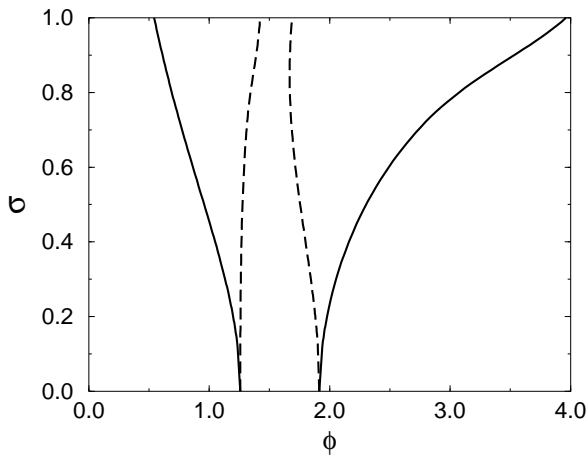,width=8.5cm}
\end{center}
\caption{ The isotropic and nematic cloud point curves (solid) and
their corresponding shadow curves (dashed) of Fig.~\protect\ref{fig1}
are replotted here, showing the (rescaled) rod volume fractions
$\mom=\m\rho$ of the coexisting phases rather than their densities
$\rho$. Of the two cloud point curves, the isotropic one is the one
with the lower $\phi$; it meets the isotropic shadow curve for
$\sigma\rightarrow0$ as it must. The nematic cloud and shadow curves
likewise coincide in this limit.
\label{fig3}
}
\end{figure}

The results for the cloud point and shadow curves are plotted in
Figs.~\ref{fig1} to~\ref{fig3}. In Fig.~\ref{fig1} we see that the
coexistence region (the density range between the isotropic and
nematic cloud points) broadens quite dramatically as the parent
distribution becomes more polydisperse. The transition, which is
already strongly first order in the monodisperse case, spreads out so
that when $\sigma=1$ the coexistence region spans almost an order of
magnitude in density, from $\rho=0.54$ to $\rho=3.96$. As $\sigma$
increases, the nematic shadow curve moves rapidly towards lower
densities, approaching the isotropic cloud curve. In Fig.~\ref{fig2a},
we show the average rod lengths in the nematic and isotropic shadow
phases. As the polydispersity increases, a strong fractionation effect
is observed, with long rods found preferentially in the nematic phase;
this is in qualitative agreement with results for the bi- and
tridisperse Onsager
model~\cite{odijk85,lekkerkerker84,vroege93,vroege97}. For $\sigma=1$,
for example, the average length of rods in the nematic phase, $\m\N$,
is more than double that in the isotropic phase, $\m\I$, both at the
isotropic and at the nematic cloud point. (Note that at the isotropic
cloud point, $\m\I=\m\pn=1$ and $\m\N>1$, while at the nematic cloud
point, $\m\N=\m\pn=1$ and $\m\I<1$.)  This fractionation effect can be
seen in more detail in Fig.~\ref{fig2b}, where for $\sigma=0.75$ we
have plotted the relevant rod length distributions $\pll$. For example,
at $l=4$ (\ie, at four times the average rod length of the parent),
$\pll$ in the nematic shadow phase is almost an order of magnitude
larger than in the isotropic cloud phase.
Finally, we study in Fig.~\ref{fig3} the cloud point and shadow curves
in a different representation~\cite{polydisp_long}: instead of the
number density $\rho$ of the coexisting phases (as in
Fig.~\ref{fig1}), we show their rescaled rod volume fraction
$\phi=\m\rho$. This leaves the cloud point curves (for which
$\m=\m\pn=1$) unchanged, but does affect the shadow curves (along
which, as Fig.~\ref{fig2a} shows, $\m$ can differ significantly from
$\m\pn=1$). Interestingly, the volume fractions of the shadow phases
turn out to depend only weakly on the polydispersity $\sigma$, in
contrast to their number densities (compare Fig.~\ref{fig1}). In fact,
the volume fraction in the nematic shadow phase does not even show a
definite trend in its dependence on polydispersity, being a
non-monotonic function of $\sigma$.

We have not so far discussed the strength of the orientational
ordering in the nematic phase. This can be characterized by the order
parameter $q=\dmom/\mom$, which has the value $q=0$ in the isotropic
phase and $q=1$ for perfect nematic ordering. It turns out that, due
to the strongly first order nature of the I-N transition, $q$ is close
to its maximal value of unity for all polydispersities $\sigma$, so we
do not display it.

\subsection{Results: Inside the coexistence region}

\begin{figure}
\begin{center}
\epsfig{file=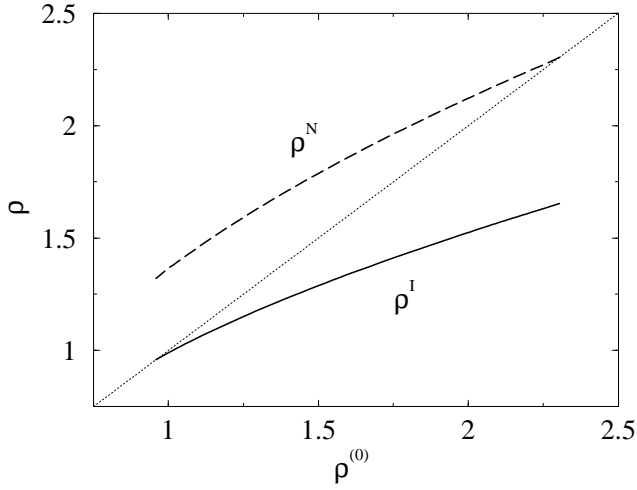,width=8.5cm}
\end{center}
\caption{The densities $\rho$ of the coexisting isotropic (solid) and
nematic (dashed) phases as a function of the parent density $\rhon$,
for polydispersity $\sigma=0.5$. The isotropic and nematic cloud
points, which delimit the coexistence region, are located at the
densities where $\rho\I$ and $\rho\N$ meet the ``dilution line''
$\rho=\rho\pn$ (dotted), respectively. Outside the coexistence region,
there is only a single isotropic (for low densities) or nematic (for
high densities) phase with density distribution $\rhonl$ (and therefore
density $\rho\pn$) identical to that of the parent.
\label{fig4}
}
\end{figure}

\begin{figure}
\begin{center}
\epsfig{file=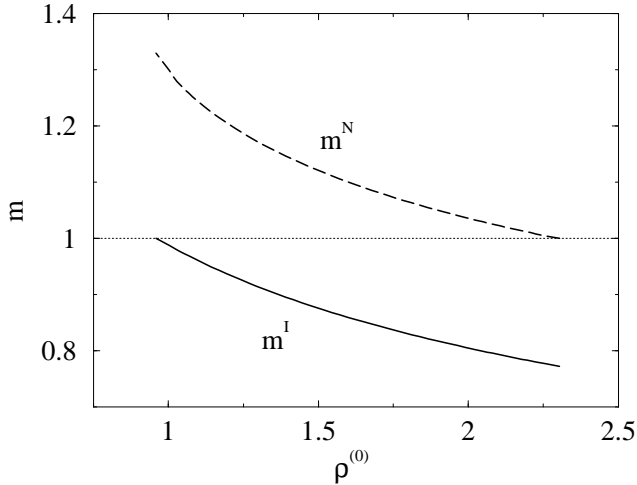,width=8.5cm}
\end{center}
\caption{ The average rod lengths in the coexisting isotropic
($\m\I$, solid) and nematic ($\m\N$, dashed) phases corresponding
to Fig.~\protect\ref{fig4}. The dotted line indicates the average
rod length $\m\pn=1$ of the parent. \label{fig5} }
\end{figure}

We now turn to the properties of the isotropic and nematic phases in
the coexistence region, \ie, for parent densities between the
isotropic and nematic cloud points. Both phases then exist in
non-infinitesimal amounts, implying $0<\gamma<1$. Using the numerical
scheme outlined in Sec.~\ref{sec:exact_coex_cond}, we then obtain the
results shown in Figs.~\ref{fig4} and~\ref{fig5}.  Fig.~\ref{fig4}
tracks the densities of the coexisting phases as the coexistence
region is crossed (for $\sigma=0.5$). As expected, the densities
interpolate between the cloud and shadow phase densities at either end
and increase smoothly with the parent density. Fig.~\ref{fig5} shows
similarly the variation of the rod lengths in the isotropic
and nematic phases across the
coexistence region. As expected from Fig.~\ref{fig2a}, the average rod
length in the nematic phase, $\m\N$, is always higher than that in the
isotropic phase, $\m\I$. At the isotropic cloud point, $\m\I=1$ and
$\m\N>1$, while at the nematic cloud point, $\m\I<1$ and $\m\N=1$;
again the values inside the coexistence region interpolate smoothly
between these limits, with both average rod lengths decreasing as the
parent density $\rhon$ increases.

Finally, we can also study the evolution of the distribution functions
$\pll$ as the fraction of volume occupied by the isotropic phase,
$\gamma$, is varied. The results are included in
Fig.~\ref{gamdist}. At the isotropic cloud point ($\gamma=1$) the
isotropic phase has the parental distribution $\pnl$; as $\gamma$ is
decreased (corresponding to increasing parent density $\rhon$), this
distribution shifts towards smaller lengths, evolving smoothly into
the distribution at the nematic cloud point $\gamma=0$. Proceeding in
the reverse direction, the nematic phase has the parent distribution
at $\gamma=0$ and then changes smoothly into the distribution at the
isotropic cloud point as $\gamma$ is increased towards $1$, shifting
towards larger rod lengths in the process.

\section{Comparison with the moment method}
\label{sec:moment}

\subsection{Constructing the moment free energy}

We now outline how the moment
method~\cite{SolCat98,Warren98,polydisp_long} can be applied to the
polydisperse Zwanzig model. To construct the moment free energy, one
recognizes from\Eq{fexc_nematic} that the excess free energy of
the model (specialized to isotropic or nematic orientational order)
only depends on the variables $\mom$ and $\dmom$. Both of these are
{\em moments} of the density distribution $\rhodl$ over lengths $l$
and orientations $d$
\beastar
\mom\equiv \rho_1 &=& \sd\intl w_1(l,d) \rhodl
\\
\dmom\equiv \rho_2 &=& \sd\intl w_2(l,d) \rhodl
\eeastar
defined by the weight functions
\bea
w_1(l,d) &=& l\nonumber\\
w_2(l,d) &=& l\left(\delta_{d,\rm z}-\frac{1}{2}\delta_{d,\rm x}
-\frac{1}{2}\delta_{d,\rm y}\right)\nonumber
\eea
We therefore call $\rho_1$ and $\rho_2$ {\em moment
densities}. Another, trivial, example of a moment density is the total
number density
\[
\rho \equiv \rho_0 = \sd \intl \rhodl
\]
which corresponds to the weight function $w_0(l,d)=1$.

Even though the {\em excess} free energy%
\[
\fexc = \frac{2}{3}(\rho_1^2-\rho_2^2)
\]
depends on moment densities only, the ideal part
\[
f_{\rm ideal} = \sd\intl\rhodl[\ln\rhodl-1]
\]
of the free energy $f=f_{\rm ideal}+\fexc$ still contains all details
of the density distribution $\rhodl$. To construct a moment free
energy which depends only on the moment densities appearing in
$\fexc$, we therefore need to transform this ideal part to a moment
form. For this purpose, it is useful to add a term $-\sd\intl \rhodl
\ln\rl = -\intl\rhol \ln \rl$ to the free energy, giving
\[
f = \sd\intl\rhodl\left[\ln\frac{\rhodl}{\rl}-1\right]+\fexc
\]
The additional term is linear in the {\em conserved} densities
$\rhol$ and therefore has no effect on the exact thermodynamics
described by $f$. (This would not be true if we had replaced $\rl$
by a $d$-dependent quantity $r_d(l)$, because $\rhodl$ is {\em
not} conserved.) For the moment method, on the other hand, $\rl$
turns out to be crucial. The key idea is to allow violations of
the particle conservation rule\eq{lever} as long as they do not
affect the moment densities appearing in the excess free energy
($\mom\equiv\rho_1$ and $\dmom\equiv\rho_2$, in our case). The
intuitive rationale---to be verified {\em a posteriori}---is that
phase behaviour is mainly governed by the excess free energy and
the moment densities appearing in it. The relevant free energy is
then obtained by minimizing $f$ at given values of the moment
densities $\rho_i$ ($i=1,2$). Using Lagrange multipliers
$\lambda_i$ to fix these values, one finds that the minimum of $f$
occurs for density distributions of the form
\be
\rhodl = \rl \exp\left[\sum_{i}\lambda_i w_i(l,d)\right]
\label{family}
\ee
and the corresponding minimum value is
\be
\fmom = \sum_{i}\lambda_i\rho_i - \rho_0 + \fexc
\label{fmom}
\ee
This expression defines the moment free energy. Note that,
from\Eq{family}, the moment densities are related to the Lagrange
multipliers by
\[
\rho_i = \sd\intl w_i(l,d)\, \rl \exp\left[\sum_{j}\lambda_j
w_j(l,d)\right]
\]
Inverting these relations determines the $\lambda_j$ in terms of the
$\rho_i$; the moment free energy\eq{fmom} thus depends on the moment
densities only, as desired.

By construction, the moment free energy\eq{fmom} is the free energy of
systems with density distributions of the form\eq{family}. One
therefore expects it to give exact results for the phase behaviour as
long as the density distributions of all the coexisting phases are
actually contained in the ``family''\eq{family}. Considering an
isotropic parent, with $\rhondl=\rhonl/3$, we can ensure that this is
true at least for the parent by choosing $\rl=\rhonl/3$. This is the
choice that we adopt from now on, giving explicitly
\be
\rhodl = \frac{1}{3}\rhonl \exp\left[\sum_{i}\lambda_i w_i(l,d)\right]
\label{pfamily}
\ee
for the family of density distributions and
\be
\rho_i = \sd\intl w_i(l,d)\, \frac{1}{3}\rhonl \exp\left[\sum_{j}\lambda_j
w_j(l,d)\right]
\label{rho_lambda}
\ee
for the relation between the Lagrange multipliers $\lambda_j$ and the
moment densities $\rho_i$. With this choice, the isotropic cloud point
and corresponding nematic shadow will be found exactly by the moment
method: at that point, the parent is only negligibly perturbed because
only an infinitesimal amount of the nematic phase has appeared, while
the nematic phase itself is related to the parent by exactly the kind
of Gibbs-Boltzmann factor appearing in\Eq{pfamily}.

To see more formally why the moment method gives exact results for
the isotropic cloud point, let us write down the resultant phase
coexistence conditions and show that they are equivalent to the
exact conditions (particle conservation violations apart).
Associated with each of the moment densities $\rho_i$ is a moment
chemical potential $\mu_i=\partial\fmom/\partial \rho_i$. Using
the Legendre transform properties~\cite{polydisp_long} of $\fmom$,
one finds
\begin{mathletters}
\label{mom_mu}
\bea
\mu_1 &=& \lambda_1 + \frac{\partial \fexc}{\partial\rho_1} = \lambda_1 +
\frac{4}{3}\rho_1
\\
\mu_2 &=& \lambda_2 + \frac{\partial \fexc}{\partial\rho_2} = \lambda_2 -
\frac{4}{3}\rho_2
\eea
\end{mathletters}
Because $\rho_1$ is conserved while $\rho_2$ is not, $\mu_1$ must be
equal in all coexisting phases, while $\mu_2$ must actually be zero:
\be
\mu_1\pa = c, \quad \mu_2\pa = 0 \qquad \mbox{for all }\al
\label{mu_eq_mom}
\ee
where $c$ is a constant common to all phases. To compare these
conditions with the exact conditions\eqq{exact_family}{lambda_cond}
for equality of the chemical potentials $\mu(l)$, we write the density
distributions\eq{pfamily} for the different rod orientations
explicitly:
\beastar
\rho_\parallel(l) \equiv \rhoz(l) &=& \frac{1}{3}\rhonl
e^{(\lambda_1+\lambda_2)l} \\
\rho_\perp(l) \equiv \rhox(l) = \rhoy(l) &=& \frac{1}{3}\rhonl
e^{(\lambda_1-\lambda_2/2)l}
\eeastar
These are of the form\eq{exact_family} if we identify
\be
\aper\equiv\lamb\x=\lamb\y=\lambda_1-\lambda_2/2,
\quad
\apar\equiv\lamb\z=\lambda_1+\lambda_2
\label{alpha_lambda}
\ee
Using\Eq{lambda_cond}, coexisting phases calculated from the moment
free energy therefore have equal (exact) chemical potentials $\mu(l)$
if
\beastar
\lambda_1\pa +\lambda_2\pa &=& 2(\mom\z\pa-\mom\pa) + c \\
&=&
\frac{4}{3}(-\mom\pa + \dmom\pa) + c \\
\lambda_1\pa - \half\lambda_2\pa &=& 2(\mom\x\pa-\mom\pa) + c \\
&=&
 \frac{2}{3}(-2\mom\pa-\dmom\pa) + c
\eeastar
in all phases. But from Eqs.\eq{mom_mu} one easily sees that these
conditions are equivalent to those [Eq.\eq{mu_eq_mom}] derived from
the moment free energy, as promised. For the condition of equality of
pressure in all phases, it is even easier to see that the moment free
energy gives the correct answer: one finds
\[
\Pi =  - \fmom + \rho_1 \mu_1 + \rho_2 \mu_2 = \rho_0 +
\frac{2}{3}(\rho_1^2-\rho_2^2)
\]
which is exactly the same as the result\eq{osm_nematic} derived from
the original free energy $f$. Remember that, in our moment density
notation, $\rho_0\equiv\rho$, $\rho_1\equiv \mom$, and
$\rho_2\equiv\dmom$.

We have thus shown that coexisting phases calculated from the moment
free energy satisfy the exact phase equilibrium conditions of equal
chemical potentials and pressures. If the phases also obey the exact
particle conservation conditions\eq{lever}, they therefore give the
exact solution of the phase coexistence problem. This is the case at
the isotropic cloud point, because one of the phases is then identical to
the isotropic parent, and the other (nematic) phase is infinitesimally
small. As stated above, this point will therefore be located exactly
by the moment free energy method. The nematic cloud point, on the
other hand, will not be found exactly: on the high-density side of
this point, the density distribution of the single nematic phase
($\rhodl=\rhonl\pdl$, with $\pdl$ obeying\Eq{pdl}), will not in
general be a member of the family\eq{pfamily}.

In Refs.~\onlinecite{SolCat98,Warren98} it was shown that the moment
free energy allows one to determine exactly the onset of phase
coexistence (cloud point and shadow), the spinodals and the critical
points of a polydisperse system with conserved densities. In our above
discussion, we have shown that for a system with non-conserved degrees
of freedom (the rod orientations), the onset of phase coexistence is
still located exactly under the following condition: in the single
phase region from which coexistence is approached, the parent must not
exhibit any ordering of the non-conserved degrees of freedoms (which
means in our case that it must be isotropic). One can show that this
conclusion holds quite generally, and that under the same restriction
spinodals and critical points found from the moment free energy also
remain exact. We thus conclude that the moment method remains useful
even in systems with non-conserved degrees of freedom.

To improve the accuracy of the moment method in the regions where it
is not exact (beyond the isotropic cloud point), one can simply retain
additional moment densities in the moment description, defined by
weight functions $w_i(l,d)$. The above construction of the moment free
energy generalizes directly to this case: the
expressions\eqqq{fmom}{pfamily}{rho_lambda} remain valid as long as
the sums over $i$ are extended appropriately. From\Eq{pfamily}, one
recognizes that the addition of new moment densities has the effect of
extending the family of density distributions that are accessible; the
exact distributions [Eq.\eq{general_rhodl}] in the coexisting phases can
thus be approximated with arbitrary accuracy as the number of moment
densities is increased. More explicitly, this can be seen as follows.
Because the additional moment densities $\rho_i$ ($i\neq 1,2$) do not
appear in the excess free energy, their associated moment chemical
potentials are simply $\mu_i = \lambda_i$. These must be equal in all
phases, so we can write the density distributions in coexisting phases
predicted by the moment method as
\beastar
\rho_d\pa(l) &=& \frac{1}{3}\rhonl \exp\left[\sum_{i\neq 1,2}\lambda_i
w_i(l,d)\right]
\\
& & \times \exp\left[\sum_{i=1,2}\lambda_i\pa
w_i(l,d)\right]
\eeastar
Comparing with\Eq{exact_family}, and bearing in mind the
identification\eq{alpha_lambda}, we see that the moment method
essentially approximates $\ln(3/D)$ (where $D$ is the denominator on
the r.h.s.\ of\Eq{general_rhodl}) by a linear combination of the
additional weight functions.  Since $D$ depends on $l$ only (not on
$d$), all these weight functions can be chosen to be $d$-independent;
the corresponding moment densities are thus conserved. A particular
additional moment density that we will always retain is the overall
density $\rho_0$, with weight function $w_0(l)=1$. This guarantees
that the dilution line $\rhol=\mbox{const}\times\rhonl
=e^{\lambda_0}\rhonl$ for the parent $\rhonl$ is contained in the
family\eq{family}, and thus simplifies the calculation of cloud points
and shadows. The optimal choice of the remaining additional weight
functions $w_i(l)$ ($i\geq 3$) is less clear cut and is discussed
further below. One thing we can say already at this point, however,
concerns the large $l$ asymptotics: it is easy to see that, for large
$l$, $\ln(3/D)=c_1 + c_2l + e^{-c_3 l}$, with constants $c_1$, $c_2$,
$c_3$, and up to terms which are exponentially smaller. The first two
contributions are covered by by the weight functions $w_0$, $w_1$,
$w_2$, so all other weight functions should be chosen to decay
exponentially for large $l$; the coefficient $c_3$ of this decay is
not known a priori, however.

\subsection{Results}

\begin{figure}[t]
\begin{center}
\epsfig{file=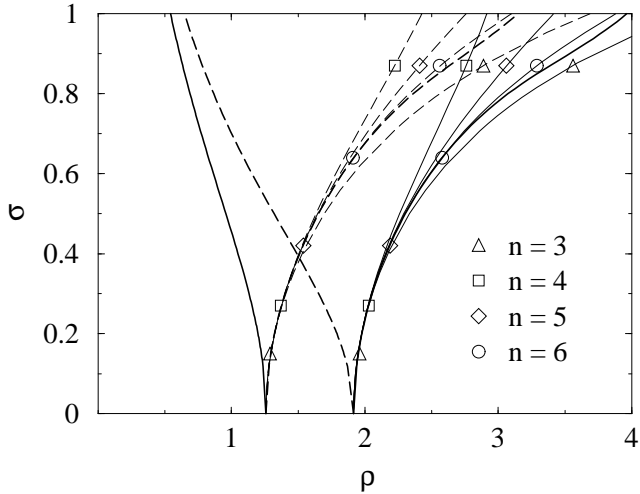,width=8.5cm}
\end{center}
\caption{Cloud point and shadow curves found using the moment
method. The exact results (compare Fig.~\protect\ref{fig1}) are
shown in bold for comparison. Even when only the minimal number of
moment densities ($n=3$) is retained in the moment free energy,
the isotropic cloud point and corresponding nematic shadow are
found exactly; the moment method results therefore overlay the
corresponding exact curves. The nematic cloud point and isotropic
shadow are not found exactly, but their accuracy increases as $n$
(indicated by the upper row of symbols) increases. The other
symbols show the points on the moment method curves where the
log-error $\delta$ first reaches the value $10^{-4}$ as $\sigma$
is increased from zero; see text for discussion.
\label{fig:mom_cp} }
\end{figure}

We show in Fig.~\ref{fig:mom_cp} the cloud point and shadow curves
obtained from the moment free energy with different numbers $n$ of
moment densities retained. As explained above we always keep, beyond
the ``essential'' moment densities $\rho_1$ and $\rho_2$, the overall
number density $\rho_0$, so the smallest value of $n$ that we consider
is $n=3$. For larger $n$, the additional weight functions were chosen
to be exponentials with increasing decay constants,
$w_{2+j}(l)=e^{-cjl}$ ($j=1\ldots n-3$).  This form is consistent with
the expected exponential behaviour for large $l$; the coefficient $c$
was chosen as $c=0.5$ by trial and error. As expected, the isotropic
cloud point curve and the corresponding nematic shadow are found
exactly already for $n=3$. For the nematic cloud point curve and the
isotropic shadow, deviations from the exact results become apparent
for larger polydispersities $\sigma$; as expected, these deviations
decrease as $n$, the number of moment densities retained, increases.

As noted in the introduction, for the polydisperse Zwanzig model
that we are considering the moment method is not really needed
because the exact phase coexistence equations can be solved
directly. However, for more realistic models (such as the
polydisperse Onsager model, with unrestricted rod orientations),
this will not be the case.  One thus needs to be able to assess
the accuracy of the moment method {\em without} knowing the exact
results beforehand. In Ref.~\onlinecite{polydisp_long}, the following
quantity was proposed for this purpose: for any phase coexistence
calculated from the moment free energy, one can work out the total
density distribution over rod lengths, $\rho_{\rm tot}(l)=\sum_\al
v\pa\rho\pa(l)=\sum_{\al,d}v\pa\rho_d\pa(l)$. The quantity
$\ln\rho_{\rm tot}(l)/\rhonl$ then measures how strongly particle
conservation of rods of length $l$ is violated; taking the square
and averaging over the normalized parent distribution
$\pnl=\rhonl/\rho\pn$ defines the ``log-error''
\be
\delta = \intl \pnl \left(\ln\frac{\rho_{\rm tot}(l)}{\rhonl}\right)^2
\label{logerr}
\ee
For small violations of particle conservation, $\ln\rho_{\rm
tot}(l)/\rhonl$ $\approx$ $\rho_{\rm tot}(l)/\rhonl - 1$, and we can think
of $\sqrt{\delta}$ as the root-mean-squared relative deviation between
$\rho_{\rm tot}(l)$ and $\rhonl$. In Fig.~\ref{fig:mom_cp}, we
indicate by the lower symbols on the nematic cloud point and isotropic
shadow curves where, as the polydispersity $\sigma$ is increased from
zero, $\delta$ first reaches the value $10^{-4}$ (which corresponds to
an average violation of particle conservation of 1\%). The fact that
the symbols lie essentially on the curves with the exact result shows
that $\delta$ provides a good indicator of the accuracy of the moment
method~\cite{polydisp_long}: adding moment densities until $\delta\leq
10^{-4}$ ensures that the results are essentially indistinguishable
from the exact ones.

\begin{figure}
\begin{center}
\epsfig{file=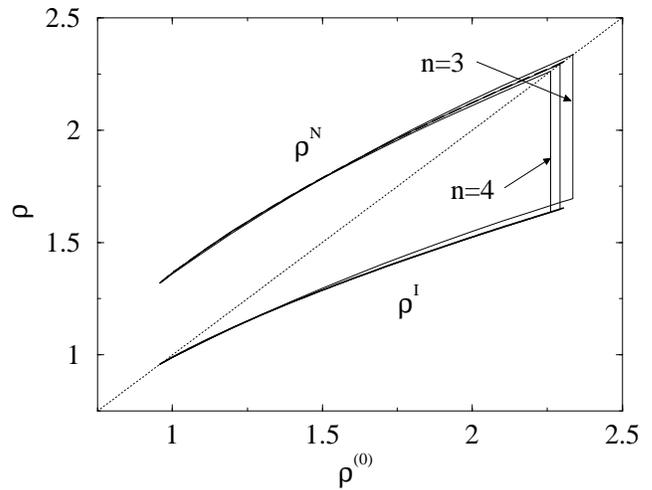,width=8.5cm}
\end{center}
\caption{The densities $\rho$ of the coexisting isotropic (solid) and
nematic (dashed) phases as a function of the parent density $\rhon$,
for polydispersity $\sigma=0.5$. We show here the results calculated
from the moment method with $n=3,4,5$ (thin lines), and the exact
results of Fig.~\protect\ref{fig4} (bold lines). As expected, the
densities beyond the isotropic cloud point are not exact, but become
increasingly more accurate as $n$ is increased. The results for $n=5$
are indistinguishable from the exact ones on the scale of the plot.
\label{fig:mom_coex}
}
\end{figure}

Beyond the calculation of phase boundaries, one would also like the
moment method to give reliable results for the properties of
coexisting phases inside the coexistence region.  In
Fig.~\ref{fig:mom_coex}, we therefore show the analogue of
Fig.~\ref{fig4} for the moment method: the densities of the coexisting
isotropic and nematic phases in the coexistence region, as a function
of the parent density $\rho\pn$. Again, exact results are obtained
only at the isotropic cloud point; but as the number of moment
densities, $n$, is increased, the results across the whole of the
coexistence region become progressively more accurate.

\begin{figure}
\begin{center}
\epsfig{file=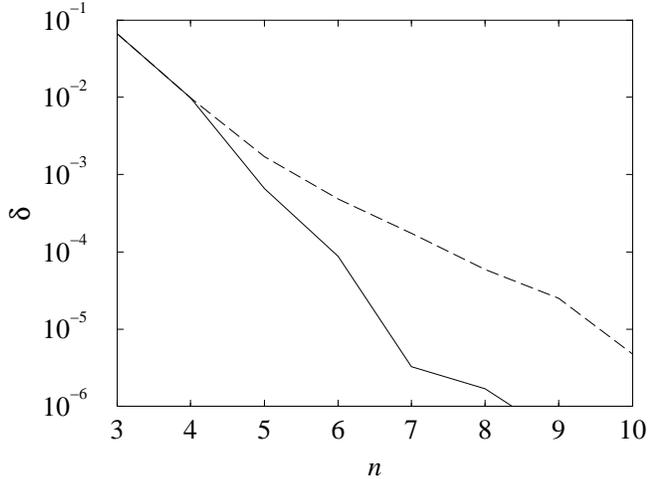,width=8.5cm}
\end{center}
\caption{The dependence of the log-error $\delta$ on the number $n$ of
moment densities retained in the moment free energy, at the point
$\rho\pn=2$, $\sigma=1$ in the phase diagram. The two lines correspond
to different choices of the additional weight functions: exponential
(solid) and power-exponential (dashed); see text for precise
functional forms.
\label{fig:delta_vs_n}
}
\end{figure}

We compare different choices for the additional weight functions in
Fig.~\ref{fig:delta_vs_n}, in terms of the dependence of the log-error
$\delta$ on $n$ at a point deep within the coexistence region of the
phase diagram. Results for two sets of additional weight functions are
shown. The first set consists of the exponential weight functions
considered above. For the second set, we chose increasing powers of
$l$, $w_{2+j}(l)=l^{j-1} e^{-cl}$ ($j=1\ldots n-3$). Note the
exponential factor, which ensures the required asymptotic behaviour;
$c$ was again chosen as 0.5. We call these weight functions
``power-exponential''.

In this example, the exponential weight functions are seen to lead to
a faster decrease of $\delta$ with $n$. However, further
experimentation with other choices of weight functions may well lead
to even better results, and it would clearly be desirable to have a
more systematic way of constructing optimal additional weight
functions. The following adaptive approach is a first step in this
direction (see also Ref.~\onlinecite{polydisp_long}). Consider a given
point in the phase diagram, characterized in the present case by the
density $\rhon$ of the parent and its polydispersity
$\sigma$. Performing a moment method phase equilibrium calculation
without additional weight functions, one will find a certain log-lever
rule violation $\ln\rho_{\rm tot}(l)/\rhonl$ (called ``log ratio'' for
short in the following) . One then expects that adding a weight
function ($w_3(l)$, in our case) which has the same $l$-dependence as
this log-ratio should significantly reduce the log-error: it extends
the family\eq{pfamily} of density distributions ``in the right
direction''. Of course, constructing $w_3(l)$ to fit the log-ratio
exactly would be computationally costly. Instead, we represent it as a
linear combination $\sum c_k\psi_k(l)$ of some simple basis functions
$\psi_k(l)$. These could be the exponential or power-exponential
functions used above, for example. The coefficients $c_k$ are chosen
to minimize the squared deviation from the log-ratio (weighted by the
normalized parent distribution),
\[
\intl \pnl \left(\sum_k c_k\psi_k(l)-
\ln\frac{\rho_{\rm tot}(l)}{\rhonl}\right)^2
\]
This is a straightforward weighted least squares problem, and the
$c_k$ can easily be found in closed form, thus determining $w_3(l)$.
One can now repeat the phase equilibrium calculation with the moment
$\rho_3$ defined by $w_3(l)$ included, and fit a new weight function
$w_4(l)$ to the resulting log-ratio (which is expected to be rather
smaller in magnitude than before). Repeating this process should lead
to a steady decrease of the log-error $\delta$. However, a large
number of additional weight functions may still be required before
$\delta$ reaches an acceptably small value, and this can cause
numerical problems~\cite{max_iterations_footnote}. To avoid this
problem, we note from the discussion at the end of the previous
section that a {\em single} additional weight function can reproduce
the exact results within the moment method, if only its $l$-dependence
can be found appropriately. Rather than keeping a large number of
additional weight functions, we can thus continually adapt a single
weight function, as follows. We choose the first additional weight
function $w_3(l)$ by fitting the initial log-ratio, rerun the phase
equilibrium calculation with $\rho_3$ included, and fit a
``temporary'' additional weight function $w_4(l)$ to the resulting
decreased log-ratio. With both $\rho_3$ and $\rho_4$ included, we
again run the calculation; this produces values of the Lagrange
multipliers $\lambda_3$ and $\lambda_4$ (which, being associated with
moment densities not appearing in the excess free energy, are common
to all phases). The key point is now that if we merge $w_3(l)$ and
$w_4(l)$ into the linear combination $w'_3(l)=\lambda_3 w_3(l) +
\lambda_4 w_4(l)$, and discard $w_4(l)$, repeating the calculation
would give exactly the same results. (All moment phase equilibrium
conditions are still satisfied, and the lever rule for $\rho'_3$
obviously follows from that for $\rho_3$ and $\rho_4$.) We are now
back to a situation with only a single additional weight function and
can repeat the process: obtain $w'_4(l)$ by fitting to the current
log-ratio, rerun with $\rho'_3$ and $\rho'_4$, combine $w'_3(l)$ and
$w'_4(l)$ into $w''_3(l)$ and so on. This method avoids the
computational problems associated with using a large number of
additional moment densities; indeed, it requires at most two
additional weight functions at any time. The number of basis
functions, however, is unrestricted in principle and can feasibly be
made quite large. In fact, one can show that for an infinitely large
set of basis functions, which allows arbitrary functional forms of the
log-ratio to be fitted, the method must converge to the results of the
exact phase equilibrium calculation (assuming it converges at
all). For finite but sufficiently large sets of basis functions, one
thus expects excellent approximations to the exact results. As long as
the set of basis functions is sufficiently ``flexible'' to approximate
the $l$-dependence of the log-ratio, the precise choice of the basis
functions should also be relatively unimportant, thus reducing the
effect of the remaining heuristic element of the method.

\begin{figure}
\begin{center}
\epsfig{file=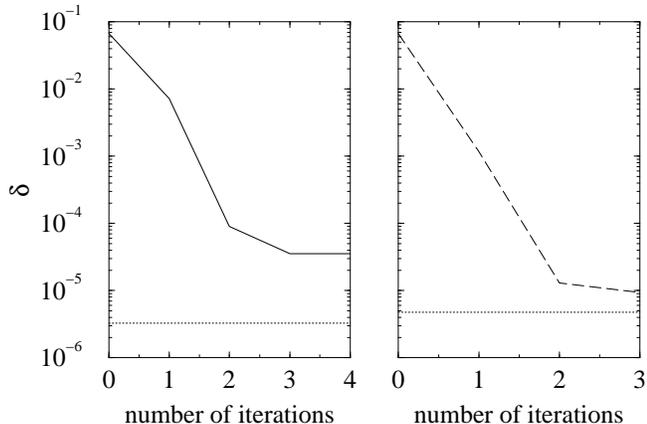,width=8.5cm}
\end{center}
\caption{The dependence of the log-error $\delta$ on the number of
iterations of the ``adaptive'' weight function algorithm described in
the text, at the point $\rho\pn=2$, $\sigma=1$ in the phase diagram.
Left: using four exponential basis functions; right: using seven
power-exponential basis functions. Dotted lines: Value of $\delta$
reached by the ``brute-force'' approach where all basis functions are
retained as weight functions.
\label{fig:delta_adaptive}
}
\end{figure}

In Fig.~\ref{fig:delta_adaptive}, we show the results for the adaptive
method just described, at the same point in the phase diagram as in
Fig.~\ref{fig:delta_vs_n}. As basis functions we considered the
exponential and power-exponential weight functions described
above. The number of basis functions was chosen such that if (in the
previous non-adaptive, ``brute-force'' approach) all basis functions
are retained as additional weight functions, the log-error $\delta$ is
less than $10^{-5}$. As can be read off from
Fig.~\ref{fig:delta_vs_n}, this leads to $n-3=7-3=4$ exponential basis
functions and $n-3=10-3=7$ power-exponential basis functions. The
corresponding ``brute-force'' values of $\delta$ are shown as
horizontal lines in Fig.~\ref{fig:delta_adaptive}. They provide
natural baselines for the results of the adaptive method: because the
latter only retains a single additional weight function (a linear
combination of the basis functions), it can obviously do no better
than the brute-force method which allows the coefficients of all basis
functions to be adjusted individually. Fig.~\ref{fig:delta_adaptive}
confirms this; the adaptive method converges after a few iterations to
a value of $\delta$ above the brute-force baseline. While the slightly
larger final value of $\delta$ is, of course, a disadvantage, the
adaptive method more than makes up for this by being much faster and
numerically more stable. We therefore plan to study this method in
more detail in future. In particular, if one is interested in
performing calculations for a number of points in the phase diagram
(along a dilution line, for example, where the density $\rhon$ of the
parent is varied) one could imagine a dynamical version of the
algorithm which adapts the single additional weight function whenever
a threshold value of the log-error $\delta$ is crossed. As long as the
chosen set of basis functions is sufficiently powerful, this should
lead to uniformly precise results across the whole phase diagram.

\begin{figure}
\begin{center}
\epsfig{file=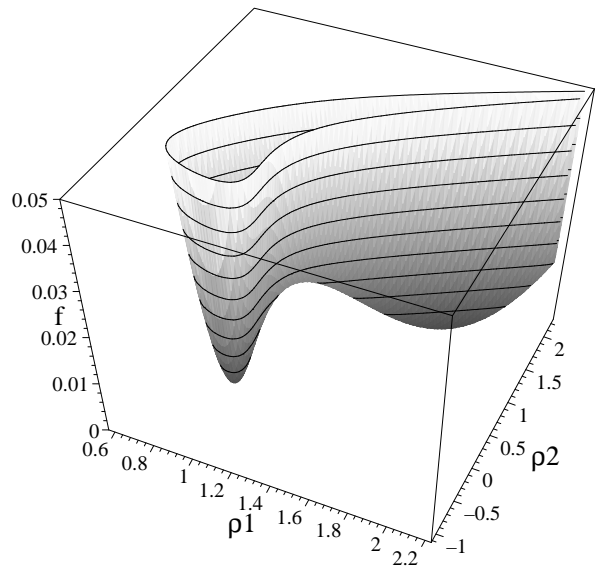,width=8cm}
\end{center}
\caption{Moment free energy $\fmom$ versus the ``essential'' moments
densities $\rho_1$ and $\rho_2$, for a Schulz parent with $\sigma=0.5$
and density $\rhon=0.959$ corresponding to the isotropic cloud
point. Constant and linear terms have been added to $\fmom$ to make
its tangent plane at the parent (represented by the point
$\rho_1=\rhon m\pn = 0.959$, $\rho_2=0$) coincide with the
$xy$-plane. As expected, this tangent plane touches the free energy
surface at a second point: the nematic shadow, whose values of
$\rho_1$ and $\rho_2$ are found exactly.
\label{fig:mom_free_en}
}
\end{figure}

Finally, we illustrate in Fig.~\ref{fig:mom_free_en} the geometrical
intuition provided by the moment free energy. We plot $\fmom$ as a
function of the ``essential'' moment densities $\rho_1$, $\rho_2$, for
a parent $\rhonl$ whose density was chosen to be exactly at the
isotropic cloud point. As expected, the tangent plane drawn at the
parent touches the surface at a second point: the nematic shadow
phase. The moment free energy thus allows a simple geometrical
interpretation of this phase transition in a polydisperse system, in
terms of a double-tangent plane to a conventional two-dimensional free
energy surface. We emphasize that the properties of the cloud and
shadow phases are found exactly, even though the moment free energy is
only a low-dimensional projection of the true free energy (which
``lives'' in the infinite-dimensional space of density distributions
$\rhodl$).

\section{Conclusion}
\label{sec:conclusion}

We have studied the phase behaviour of the Zwanzig model of
suspensions of hard rods, allowing for polydispersity in the lengths
of the rods. The model assumes that the rods are restricted to lie
along one of three orthogonal axes. In spite of this drastic
simplification (compared to the Onsager model, where rod orientations
are unrestricted), the results we obtain are in qualitative agreement
with experimental observations~\cite{buining93,vanbruggen96}: the
coexistence region broadens significantly as the polydispersity (the
width of the rod length distribution of the parent) increases;
fractionation is also observed, with long rods found preferentially in
the nematic phase. These conclusions were obtained from an exact
analysis of the phase equilibrium equations, starting from the free
energy of the model within the second virial approximation.

In the second part of the paper, we considered the application of
the moment method to the polydisperse Zwanzig model. This involved
extending the construction of the moment free energy to models
with both conserved and non-conserved degrees of freedom. We
showed that most of the exactness statements obtained previously
for systems with conserved densities carry over to this case: the
onset of phase coexistence is still found exactly from the moment
free energy, as long as it is approached from a single phase
region where there is no ordering of the non-conserved degrees of
freedom. With the same restriction, spinodal instabilities and
critical points are also located exactly. Our concrete results for
the cloud and shadow curves bear this out; the isotropic cloud
point and corresponding nematic shadow are found exactly, while
the nematic cloud point and isotropic shadow are approximate. The
accuracy of the approximation increases as the number of moment
densities retained in the moment free energy is increased. The
log-error $\delta$ is a useful criterion for monitoring the
increase in accuracy; crucially, it can be computed without
knowing the exact results beforehand. Finally, we have discussed
methods for choosing the weight functions of the additional moment
densities. An adaptive technique, which requires at most two
additional weight functions at any given time, and is therefore
very cheap to implement computationally, gives promising results.

In future work, it is obviously desirable to remove the
simplications of the Zwanzig model and move towards a study of the
polydisperse Onsager model, with unrestricted rod orientations. A
direct numerical solution of the phase equilibrium equations for
this model is infeasible, so an approach based on the moment
method suggests itself. One complication is that the excess free
energy does not have a simple moment structure: the excluded
volume between two rods at angles $\theta$ and $\theta'$ with the
nematic axis is a non-trivial function of these angles. Its
expansion in terms of Legendre polynomials~\cite{KayRav78} shows
that the moment free energy actually depends on an infinite number
of moment densities. As an intermediate step, we therefore plan to
consider the polydisperse Maier-Saupe model~\cite{MaiSau58}, which
truncates the eigenfunction expansion after the leading term and
leads to an excess free energy depending on only two moment
densities. This approach should be of independent interest as a
phenomenological description of polydisperse suspensions of
rod-like particles with more complex (soft) interactions. Finally,
it would also be interesting to study the effect of
diameter-polydispersity on hard rod systems. Novel features such
as isotropic-isotropic phase coexistence have previously been
found for the bidisperse case (rods with two different
diameters)~\cite{SeaJac95,SeaMul96,VanMul96b,VanMulDij98}, and it will
be interesting to see how these are modified for truly
polydisperse systems.

{\bf Acknowledgments:} JC's work is part of the project PB96-0119
of the Direc\-ci\'{o}n General de Ense\-\~{n}an\-za Superior. JC and
RS are partly supported by project HB1998-0008 funded by Acciones
Integradas con el Reino Unido (Ministerio de Educaci{\'{o}}n y
Cultura) and The British Council. Support from the EPSRC Soft
Condensed Matter Network is gratefully acknowledged.


\begin{thebibliography}{10}
\bibitem{max_iterations_footnote}
Note that once the number of additional weight functions equals the number of
  basis functions, the iterative process can be stopped because no further
  improvement is possible: the adaptively chosen weight functions then span the
  same space as the basis functions. All results will therefore be the same as
  if the basis functions themselves had been chosen as the additional weight
  functions, and adding further weight functions does not enlarge the
  family~\protect(\ref{pfamily}) of accessible density distributions.
\end{thebibliography}

\end{document}